\documentclass[aps,physrev,reprint,groupedaddress]{revtex4-2}
\usepackage{xspace,xcolor,graphicx,gensymb,amsmath,multirow,longtable,xtab,afterpage,lipsum,listings,comment,verbatim,blindtext,float}
\providecommand{\units}[1]{\,\ensuremath{\mathrm{#1}} \xspace}

\begin{document}

\title{Determination of the speed of light from the longitudinal modes beat frequencies of an open-cavity helium-neon laser}
\author{Mohamad Taim}
\thanks{Communicating author}
\email{taim0001@umn.edu}
\author{Nicholas Kuder}
\email{kuder023@umn.edu}

\affiliation{University of Minnesota, Minneapolis, MN, USA}

\date{\today}

\begin{abstract}
    We empirically determined the speed of light by measuring the variation in longitudinal mode frequencies, or the beat frequencies, of an adjustable-length, open-cavity helium-neon laser as a function of its cavity length. The TEM$_{00}$ mode lasing output of the laser was analyzed using a fast frequency photodiode detector and a radio frequency spectrum analyzer. A Fabry-Perot interferometer was used to monitor the intensity of the longitudinal modes and we found that the phenomena of frequency pushing and pulling had little effects on the beat frequency measurements. Plotting the reciprocal of the beat frequency as a function of the change in cavity length, the speed of light was found, by using linear weighted least squares regression, to be $(2.997 \pm 0.003) \times 10^{8} \units{m s^{-1}}$. This value is $0.3 \sigma$ away from the defined value of speed of light and is accurate to 1 part in 3200.
\end{abstract}

\maketitle
\section{Introduction}
    The speed of light, $c$, is an important fundamental constant in many scientific fields, from electrodynamics to general relativity. $c$ is not just about describing the propagation property of electromagnetic waves, but it also happens to be the upper bound to the propagation speed of signals and to the speeds of all material particles. 
    
    Large numbers of determinations have been made since Galilei Galileo first attempted to prove the quantifiability of speed of light by using two lanterns placed on top of two separate hills and measuring the time taken for the light signals to cross the valley \cite{galileo}. Most contemporary attempts on measuring the speed of light employed variations of the two common experimental methods: time of flight, and measurement of the frequency and wavelength of a radiation \cite{towards}. Earlier papers described techniques like Foucalt’s rotating mirror method \cite{foucalt}, Kerr electro-optical shutter \cite{kerr, kerr2}, and microwaves \cite{microwave}.
    
    Our experiment utilized the resonance condition of an open cavity, first done by Essen and Gordon-Smith in 1947 with a cylindrical copper cavity resonator as described in Ref. \cite{cylindrical}. Their value of $c$ was accurate to 3 parts in 100,000 In 2010, the determination in Ref. \cite{d'orazio} utilized a Helium-Neon (He-Ne) laser to make a determination that was accurate to within 1 in 4100. We also used a He-Ne laser system to determine $c$ by studying the longitudinal mode frequencies of the lasing beam and its relationship with the length of the laser cavity.

\section{Theory}
    A He-Ne laser system consists of a cavity in between two highly reflective mirrors, or a Fabry-Perot cavity. A 99.9\% reflective mirror is on one end and a 99\% reflective output coupler is on the other with the output coupler letting through a little amount of light that becomes the laser beam. The cavity is filled with a mixture of helium and neon and together, the gas-filled cavity is called the medium.
    
    When a DC current is pumped into the medium, the electrons in it excite into higher levels of energy and the medium is said to undergo population inversion, a process in which the number of excited atoms is more than the number of ground atoms. These excited electrons then decay to lower levels of energy and release photons. This is the spontaneous emission and the photons are emitted in random directions and phases. Those that are parallel to the horizontal axis of the laser tube get reflected by the two mirrors many, many times. This leads to stimulated emission when the photons interact with other excited electrons in the medium. The result is an avalanche of photons with similar orientation, wavelength, and phase that comes out of the output coupler as a bright, nearly monochromatic, single beam of laser.
    
    The Fabry-perot cavity also imposes a resonance condition on the system in that the laser will not operate with just any wavelength. The lasing output will only have wavelengths, $\lambda$, that satisfy the equation
    \begin{equation}
        \label{eq:lambda}
        \lambda = \frac{2nL}{k} \text{,}
    \end{equation}
    or, converted into mode frequency, $f$, in the equation
    \begin{equation}
        \label{eq:f}
        f = \frac{kc}{2nL} \text{,}
    \end{equation}
    where $n$ is the index of refraction of the medium in which the light wave travels, $L$ is the length of the cavity, and $k$ is an integer representing the mode number \cite{d'orazio}.
    
\begin{figure}[H]
        \includegraphics[width = \linewidth]{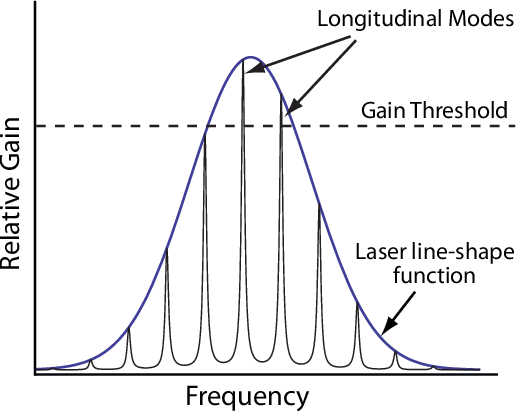}
        \caption{The modes of a laser output under the medium gain curve.}
        \label{gauss}
    \end{figure}

    As shown in Fig. \ref{gauss}, a laser beam has a Gaussian-shaped output \cite{d'orazio}. This is due to Doppler broadening, a spectrum of Doppler shifting effects caused by the random directions of the photons in the medium.
    Combined with the fact that the medium has a gain threshold, we can only see two or three mode frequencies at a given time, and the difference between any two of these mode frequencies is known as the beat frequency, $f_b$, given as
    \begin{equation}
        \label{eq:fb}
        f_b = \frac{c}{2nL} \text{.}
    \end{equation}
    From this equation, assuming a fixed value for $n$, we can know $c$ accurately by measuring and plotting only the two variables, $f_b$ and $L$.
    
\section{Experimental Setup \& Methodology}
    \subsection{Laboratory Setup}
        A He-Ne laser tube is mounted on an optical post and a $99.9 \units{\%}$ reflective output coupler (ThorLabs) with a radius of $45 \units{cm}$ is secured on a translation stage (Micro-Controle), which is controlled by a stepper motor (PhobOTronix), completing the open cavity of the laser system. Between the output coupler and the laser tube is an adjustable iris to block unwanted transverse modes.
    
        The laser beam travels through a non-polarizing beam splitter, which directs half of the beam to a fast frequency photodiode detector (Thorlabs model PDA8A) and the other half into a scanning Fabry-Perot interferometer (SFPI) (Tropel model 240). The signal from the photodetector is sent to a radio frequency (RF) spectrum analyzer (Rigol model DSA815), which displays the beat frequency. The beam from the SFPI is directed to another photodetector, which is connected to an oscilloscope that will display the relative intensities of the modes. Both the spectrum analyzer and the SFPI are assumed to be calibrated correctly.
        
        The measurements on beat frequency will be affected by two optical phenomena, namely frequency pulling and pushing. Frequency pulling occurs when there are varying index of refraction near the resonance transition. This causes the difference in mode frequencies, depending on its position under the gain profile curve, to have a smaller value \cite{fpulling}. In other words, the two mode frequencies are being pulled toward the center. This effect is more pronounced when we have mode frequencies that are not at similar intensity.
        
        Meanwhile, frequency pushing is when we have an increase in gain, or increase in refractive index, due to reasons that are myriad and complex, such as the ambient temperature and pressure. This phenomenon increases the beat frequency and also the total intensity \cite{fpulling}. We can prevent this by taking data only when the total intensity is not changing.
        
        The SFPI serves to monitor the relative intensities of the longitudinal modes. The iris in between the Brewster window and the output coupler is used to adjust the total intensity of the laser beam. These measures are taken to minimize the uncertainty in beat frequency measurements.
        
        The schematic diagram of the experiment is shown below in Fig. \ref{schematic} \cite{d'orazio}.
        \begin{figure}[h!]
            \includegraphics[width = \linewidth]{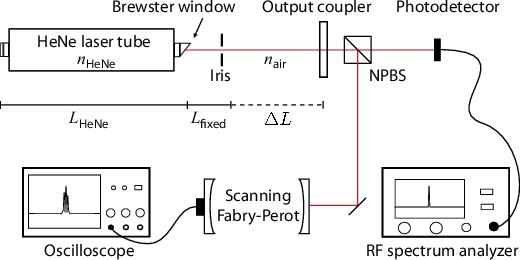}
            \caption{The schematic diagram of the experimental setup.}
            \label{schematic}
        \end{figure}
        
    \subsection{Calibration}
        The stepper motor, controlled via a LabVIEW (version 2018) program, is used to move the translation stage by a known number of steps. After the translation stage translated the distance, a digital caliper, fixed on the translation stage using tapes to minimize human error, is used to measure the total distance moved by the translation stage. This is the change in laser cavity length, $\Delta L$. This process is repeated until enough data points have been collected. The change in length per step is then determined by the slope of the data using linear regression.
    
\section{Results \& Data Analysis}
\begin{figure}[h!]
            \includegraphics[width = \linewidth]{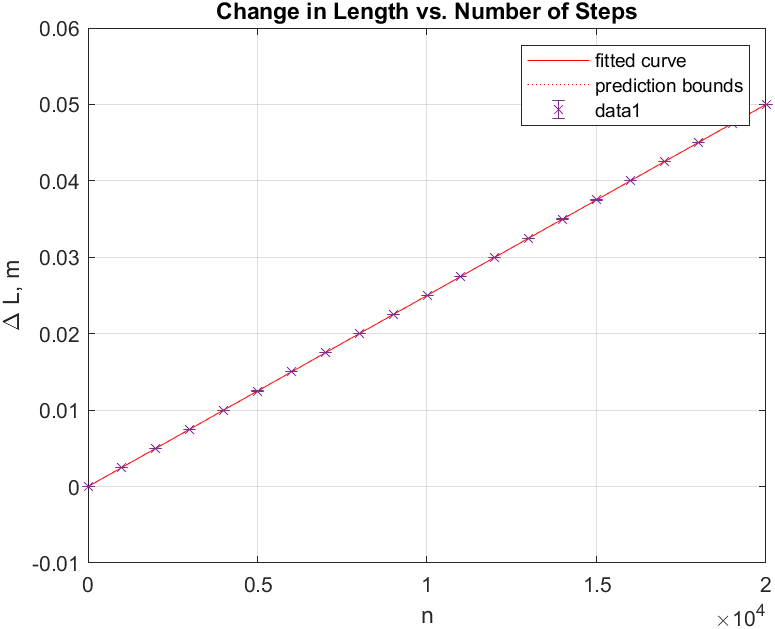}
            \caption{The plot for the calibration of the translation stage.}
            \label{calibration}
    \end{figure}

    In order to calibrate the translation stage and obtain a relationship between the number of steps, $n$, and the translated distance, $\Delta L$, which corresponds to the difference in length of the laser cavity, we measured $\Delta L$ for a known value of $n$. We found from the plot in Fig. \ref{calibration}, for $n = 10$,
    \begin{equation}
        \Delta L = 2.5 \times 10^{-5} \units{m}
    \end{equation}
    The uncertainty in $\Delta L$ is $\sigma_{\Delta L} = \pm 1.0 \times 10^{-5}$, which is dictated by the tolerance of the digital caliper.
    
\begin{figure}[h!]
        \includegraphics[width = \linewidth]{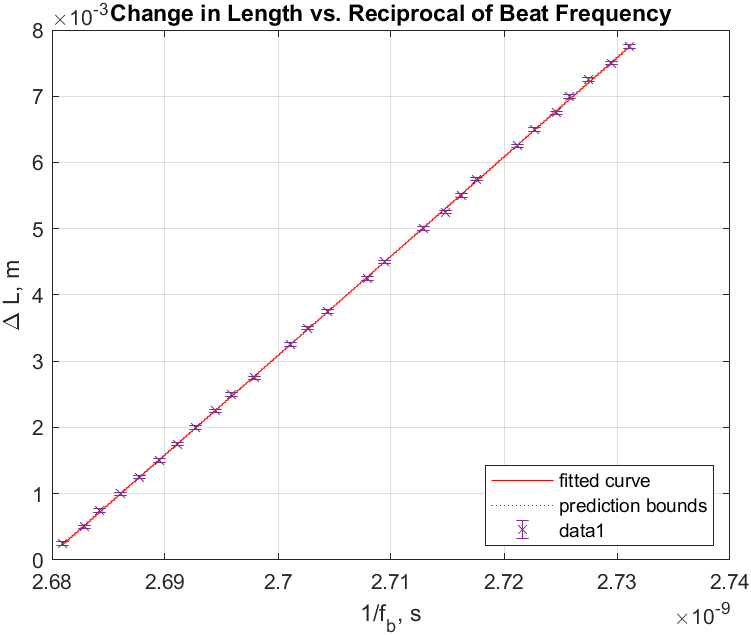}
        \caption{Reciprocal beat frequency as a function of the change in the open-cavity length.}
        \label{plot}
    \end{figure}

    Using the iris to ensure that the mode is in TEM$_{00}$ mode (single circular beam), we then measured the beat frequencies as a result of varying the length of the laser cavity. The values for the reciprocal of beat frequency, $1/f_b$, as a function of $\Delta L$ is plotted in Fig. \ref{plot}.
    
    For each $\Delta L$, the value for $1/f_b$ was averaged over 100 sweep cycles to minimize the uncertainty on frequency reading by an order of magnitude.
    
\begin{figure}[h!]
        \includegraphics[width = \linewidth]{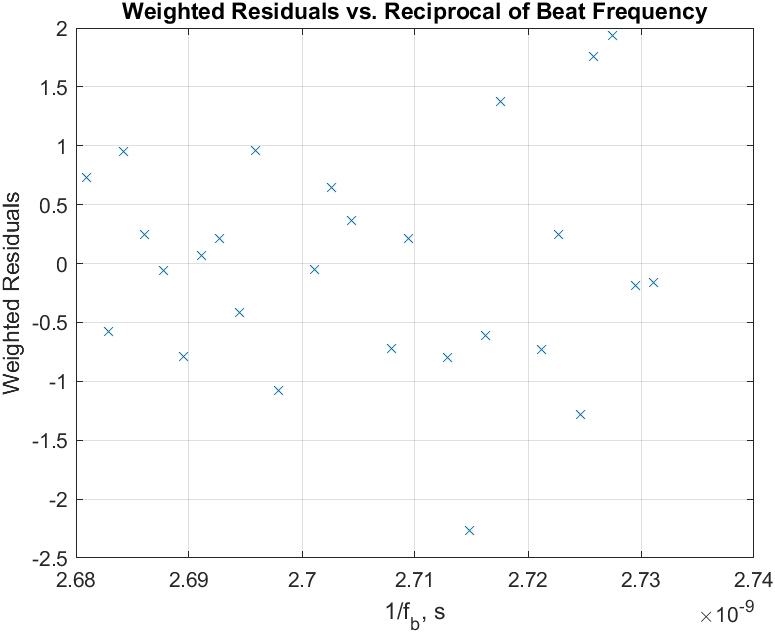}
        \caption{Residuals from the linear fit.}
        \label{residuals}
    \end{figure}

    Linear regression using the weighted least squares method was performed and the data plotted in Fig. \ref{plot} yielded the final result for the speed of light in air to be
    \begin{equation}
        c = (2.997 \pm 0.003) \times 10^8 \units{m s^{-1}}
    \end{equation}
    Our measured value is accurate to within $0.3 \sigma$ of the defined value $c = 299,792,458 \units{m s^{-1}}$ \cite{codata}. As can be seen in Fig. \ref{residuals}, the linear fit residuals show no structure, which indicates good data. Specifically, the reduced chi squared is $\chi^2_{\mu} = 0.922$ and the P-value is $P = 0.58$.
    
    The uncertainty in the final result can be explained by the phenomena frequency pushing and pulling. Although we found no correlation between the two phenomena and the beat frequency. Specifically, we found that the relative intensities of the two modes observed on the SFPI oscilloscope had little affect on the value of beat frequency on the spectrum analyzer. Further investigation also found little effect on beat frequency caused by the total intensity, which we controlled by the iris.  However, we managed to get good data in the sense that we measured the beat frequency only when the total intensity and the relative intensity of the two modes are similar from one measurement to the next.
    
    In future experiments, one can reduce the uncertainty caused by frequency pulling by implementing an electronic feedback system that can make slight adjustments to the cavity length in order to maintain similar relative intensities at all time. Other sources of error like the misalignment of the optical equipment, and the thermal expansion of the optical table can also be looked into.

\section{Conclusions}
    by varying the length of the open cavity of a He-Ne laser, we determined the speed of light to be $c = (2.997 \pm 0.003) \times 10^8 \units{m s^{-1}}$. This value is $0.3 \sigma$ away from the defined value $c = 299,792,458 \units{m s^{-1}}$. We found that the phenomena frequency pulling and pushing had little to no effects on our measurements and thus, negligible uncertainties on our final value of $c$.


%

\begin{acknowledgments}
    Our thanks to Professor Kurt Wick for his guidance and enthusiasm throughout the research; to Professors Dan Dahlberg and Daniel Cronin-Hennesy for their lectures; to Kevin Booth for the constant help with various problems. We express gratitute to the Department of Physics for the equipments and the permission to run the experiment.
\end{acknowledgments}

\nocite*
\bibliography{refs}

\providecommand{\noopsort}[1]{}\providecommand{\singleletter}[1]{#1}
\begin{thebibliography}{10}%
\makeatletter
\providecommand \@ifxundefined [1]{%
 \@ifx{#1\undefined}
}%
\providecommand \@ifnum [1]{%
 \ifnum #1\expandafter \@firstoftwo
 \else \expandafter \@secondoftwo
 \fi
}%
\providecommand \@ifx [1]{%
 \ifx #1\expandafter \@firstoftwo
 \else \expandafter \@secondoftwo
 \fi
}%
\providecommand \natexlab [1]{#1}%
\providecommand \enquote  [1]{``#1''}%
\providecommand \bibnamefont  [1]{#1}%
\providecommand \bibfnamefont [1]{#1}%
\providecommand \citenamefont [1]{#1}%
\providecommand \href@noop [0]{\@secondoftwo}%
\providecommand \href [0]{\begingroup \@sanitize@url \@href}%
\providecommand \@href[1]{\@@startlink{#1}\@@href}%
\providecommand \@@href[1]{\endgroup#1\@@endlink}%
\providecommand \@sanitize@url [0]{\catcode `\\12\catcode `\$12\catcode
  `\&12\catcode `\#12\catcode `\^12\catcode `\_12\catcode `\%12\relax}%
\providecommand \@@startlink[1]{}%
\providecommand \@@endlink[0]{}%
\providecommand \url  [0]{\begingroup\@sanitize@url \@url }%
\providecommand \@url [1]{\endgroup\@href {#1}{\urlprefix }}%
\providecommand \urlprefix  [0]{URL }%
\providecommand \Eprint [0]{\href }%
\providecommand \doibase [0]{https://doi.org/}%
\providecommand \selectlanguage [0]{\@gobble}%
\providecommand \bibinfo  [0]{\@secondoftwo}%
\providecommand \bibfield  [0]{\@secondoftwo}%
\providecommand \translation [1]{[#1]}%
\providecommand \BibitemOpen [0]{}%
\providecommand \bibitemStop [0]{}%
\providecommand \bibitemNoStop [0]{.\EOS\space}%
\providecommand \EOS [0]{\spacefactor3000\relax}%
\providecommand \BibitemShut  [1]{\csname bibitem#1\endcsname}%
\let\auto@bib@innerbib\@empty
\bibitem [{\citenamefont {Foschi}\ and\ \citenamefont {Leone}(2009)}]{galileo}%
  \BibitemOpen
  \bibfield  {author} {\bibinfo {author} {\bibfnamefont {R.}~\bibnamefont
  {Foschi}}\ and\ \bibinfo {author} {\bibfnamefont {M.}~\bibnamefont {Leone}},\
  }\bibfield  {title} {\bibinfo {title} {Measurement of the velocity of light,
  and the reaction times},\ }\href@noop {} {\bibfield  {journal} {\bibinfo
  {journal} {Perception}\ }\textbf {\bibinfo {volume} {38}},\ \bibinfo {pages}
  {1251} (\bibinfo {year} {2009})}\BibitemShut {NoStop}%
\bibitem [{\citenamefont {Bradley}\ \emph {et~al.}(1972)\citenamefont
  {Bradley}, \citenamefont {Edwards}, \citenamefont {Knight}, \citenamefont
  {Rowley},\ and\ \citenamefont {Woods}}]{towards}%
  \BibitemOpen
  \bibfield  {author} {\bibinfo {author} {\bibfnamefont {C.~C.}\ \bibnamefont
  {Bradley}}, \bibinfo {author} {\bibfnamefont {G.~J.}\ \bibnamefont
  {Edwards}}, \bibinfo {author} {\bibfnamefont {D.~J.~E.}\ \bibnamefont
  {Knight}}, \bibinfo {author} {\bibfnamefont {W.~R.~C.}\ \bibnamefont
  {Rowley}},\ and\ \bibinfo {author} {\bibfnamefont {P.~T.}\ \bibnamefont
  {Woods}},\ }\bibfield  {title} {\bibinfo {title} {Towards a new determination
  of the speed of light},\ }\href@noop {} {\bibfield  {journal} {\bibinfo
  {journal} {Phys. Bull.}\ }\textbf {\bibinfo {volume} {23}},\ \bibinfo {pages}
  {15} (\bibinfo {year} {1972})}\BibitemShut {NoStop}%
\bibitem [{\citenamefont {Domkowski}\ \emph {et~al.}(1972)\citenamefont
  {Domkowski}, \citenamefont {Richardson},\ and\ \citenamefont
  {Rowbotham}}]{foucalt}%
  \BibitemOpen
  \bibfield  {author} {\bibinfo {author} {\bibfnamefont {A.~J.}\ \bibnamefont
  {Domkowski}}, \bibinfo {author} {\bibfnamefont {C.~B.}\ \bibnamefont
  {Richardson}},\ and\ \bibinfo {author} {\bibfnamefont {N.}~\bibnamefont
  {Rowbotham}},\ }\bibfield  {title} {\bibinfo {title} {Measurement of the
  speed of light},\ }\href@noop {} {\bibfield  {journal} {\bibinfo  {journal}
  {Am. J. Phys.}\ }\textbf {\bibinfo {volume} {40}},\ \bibinfo {pages} {910}
  (\bibinfo {year} {1972})}\BibitemShut {NoStop}%
\bibitem [{\citenamefont {Palmer}\ and\ \citenamefont {Spratt}(1954)}]{kerr}%
  \BibitemOpen
  \bibfield  {author} {\bibinfo {author} {\bibfnamefont {C.~H.}\ \bibnamefont
  {Palmer}}\ and\ \bibinfo {author} {\bibfnamefont {G.~S.}\ \bibnamefont
  {Spratt}},\ }\bibfield  {title} {\bibinfo {title} {A laboratory experiment on
  the velocity of light},\ }\href@noop {} {\bibfield  {journal} {\bibinfo
  {journal} {Am. J. Phys.}\ }\textbf {\bibinfo {volume} {22}},\ \bibinfo
  {pages} {481} (\bibinfo {year} {1954})}\BibitemShut {NoStop}%
\bibitem [{\citenamefont {Rogers}\ \emph {et~al.}(1969)\citenamefont {Rogers},
  \citenamefont {McMillan}, \citenamefont {Pickett},\ and\ \citenamefont
  {Anderson}}]{kerr2}%
  \BibitemOpen
  \bibfield  {author} {\bibinfo {author} {\bibfnamefont {J.}~\bibnamefont
  {Rogers}}, \bibinfo {author} {\bibfnamefont {R.}~\bibnamefont {McMillan}},
  \bibinfo {author} {\bibfnamefont {R.}~\bibnamefont {Pickett}},\ and\ \bibinfo
  {author} {\bibfnamefont {R.}~\bibnamefont {Anderson}},\ }\bibfield  {title}
  {\bibinfo {title} {A determination of the speed of light by the phase-shift
  method},\ }\href@noop {} {\bibfield  {journal} {\bibinfo  {journal} {Am. J.
  Phys.}\ }\textbf {\bibinfo {volume} {37}},\ \bibinfo {pages} {816} (\bibinfo
  {year} {1969})}\BibitemShut {NoStop}%
\bibitem [{\citenamefont {Bol}(1950)}]{microwave}%
  \BibitemOpen
  \bibfield  {author} {\bibinfo {author} {\bibfnamefont {K.}~\bibnamefont
  {Bol}},\ }\bibfield  {title} {\bibinfo {title} {A determination of the speed
  of light by the resonant cavity method},\ }\href@noop {} {\bibfield
  {journal} {\bibinfo  {journal} {Phys. Rev.}\ }\textbf {\bibinfo {volume}
  {80}},\ \bibinfo {pages} {298} (\bibinfo {year} {1950})}\BibitemShut
  {NoStop}%
\bibitem [{\citenamefont {Essen}\ \emph {et~al.}(1948)\citenamefont {Essen},
  \citenamefont {Gordon-Smith},\ and\ \citenamefont {Darwin}}]{cylindrical}%
  \BibitemOpen
  \bibfield  {author} {\bibinfo {author} {\bibfnamefont {L.}~\bibnamefont
  {Essen}}, \bibinfo {author} {\bibfnamefont {A.~C.}\ \bibnamefont
  {Gordon-Smith}},\ and\ \bibinfo {author} {\bibfnamefont {C.~G.}\ \bibnamefont
  {Darwin}},\ }\bibfield  {title} {\bibinfo {title} {The velocity of
  propagation of electromagnetic waves derived from the resonant frequencies of
  a cylindrical cavity resonator},\ }\href@noop {} {\bibfield  {journal}
  {\bibinfo  {journal} {Proc. R. Soc. A}\ }\textbf {\bibinfo {volume} {194}},\
  \bibinfo {pages} {348} (\bibinfo {year} {1948})}\BibitemShut {NoStop}%
\bibitem [{\citenamefont {D’Orazio}\ \emph {et~al.}(2010)\citenamefont
  {D’Orazio}, \citenamefont {Pearson}, \citenamefont {J.~T.~Schultz},
  \citenamefont {Best}, \citenamefont {Goodfellow}, \citenamefont {Scholten},\
  and\ \citenamefont {White}}]{d'orazio}%
  \BibitemOpen
  \bibfield  {author} {\bibinfo {author} {\bibfnamefont {D.~J.}\ \bibnamefont
  {D’Orazio}}, \bibinfo {author} {\bibfnamefont {M.~J.}\ \bibnamefont
  {Pearson}}, \bibinfo {author} {\bibfnamefont {D.~S.}\ \bibnamefont
  {J.~T.~Schultz}}, \bibinfo {author} {\bibfnamefont {M.~W.}\ \bibnamefont
  {Best}}, \bibinfo {author} {\bibfnamefont {K.~M.}\ \bibnamefont
  {Goodfellow}}, \bibinfo {author} {\bibfnamefont {R.~E.}\ \bibnamefont
  {Scholten}},\ and\ \bibinfo {author} {\bibfnamefont {J.~D.}\ \bibnamefont
  {White}},\ }\bibfield  {title} {\bibinfo {title} {Measuring the speed of
  light using beating longitudinal modes in an open-cavity {HeNe} laser},\
  }\href@noop {} {\bibfield  {journal} {\bibinfo  {journal} {Am. J. Phys.}\
  }\textbf {\bibinfo {volume} {78}},\ \bibinfo {pages} {524} (\bibinfo {year}
  {2010})}\BibitemShut {NoStop}%
\bibitem [{\citenamefont {Lindberg}(1999)}]{fpulling}%
  \BibitemOpen
  \bibfield  {author} {\bibinfo {author} {\bibfnamefont {{\AA}.~M.}\
  \bibnamefont {Lindberg}},\ }\bibfield  {title} {\bibinfo {title} {Mode
  frequency pulling in {He–Ne} lasers},\ }\href@noop {} {\bibfield  {journal}
  {\bibinfo  {journal} {Am. J. Phys.}\ }\textbf {\bibinfo {volume} {67}},\
  \bibinfo {pages} {350} (\bibinfo {year} {1999})}\BibitemShut {NoStop}%
\bibitem [{\citenamefont {Tiesinga}\ \emph {et~al.}()\citenamefont {Tiesinga},
  \citenamefont {Mohr}, \citenamefont {Newell},\ and\ \citenamefont
  {Taylor}}]{codata}%
  \BibitemOpen
  \bibfield  {author} {\bibinfo {author} {\bibfnamefont {E.}~\bibnamefont
  {Tiesinga}}, \bibinfo {author} {\bibfnamefont {P.~J.}\ \bibnamefont {Mohr}},
  \bibinfo {author} {\bibfnamefont {D.~B.}\ \bibnamefont {Newell}},\ and\
  \bibinfo {author} {\bibfnamefont {B.~N.}\ \bibnamefont {Taylor}},\
  }\href@noop {} {\bibinfo {title} {The 2018 {CODATA} recommended values of the
  fundamental physical constants}},\ \bibinfo {howpublished}
  {\url{https://physics.nist.gov/cgi-bin/cuu/Value?c}}\BibitemShut {NoStop}%
\end{thebibliography}%
\end{document}